%Paper: astro-ph/9401051
%From: Ray Carlberg <carlberg@hermes.astro.washington.edu>
%Date: Mon, 31 Jan 1994 09:35:41 -0800 (PST)

\newif\ifpics
\newif\ifpairpics
%%% UNCOMMENT TO GET PICTURES ...
\picstrue
\newif\ifapj
%%%%\apjtrue
\ifapj
	\magnification=\magstep1
\else
	\magnification=\magstephalf
\fi

\input epsf
\epsfverbosetrue
\font\caps=cmcsc10

%\raggedbottom     %Keeps spaces in document from stretching.%

\ifapj\baselineskip 24pt \else \baselineskip=15truept \fi
\overfullrule 0pt
\parindent=2.5em        %Sets paragraph indent to 5 spaces.%
\def\ub{\underbar}
\def\hi{\par\noindent \hangindent=2.5em}
\def\ls{\vskip 12.045pt}   %One line space.%
\def\ni{\noindent}        %No Indent%
\def\et{{\it et\thinspace al.}\ }    %et al.%
\def\eg{{\it e.\thinspace g.}\ }    %e.g.%
\def\ie{{\it i.\thinspace e.}\ }    %i.e.%
\def\kms{km\thinspace s$^{-1}$ }     %kms -1%
     %A mm-1%
\def\deg{\ifmmode^\circ\else$^\circ$\fi}    %Degree sign%

\def\arcs{\ifmmode {'' }\else $'' $\fi}     %Arc seconds%
\def\arcm{\ifmmode {' }\else $' $\fi}     %Arc minutes%
\def\buildrel#1\over#2{\mathrel{\mathop{\null#2}\limits^{#1}}}
\def\mper{\ifmmode \buildrel m\over . \else $\buildrel m\over .$\fi}
%Superscript 'm' over period.%
\def\hper{\ifmmode \rlap.^{h}\else $\rlap{.}^h$\fi}
%Superscript 'h' over period%
\def\sper{\ifmmode \rlap.^{s}\else $\rlap{.}^s$\fi}
%Superscript 's' over period.%
\def\arcsper{\ifmmode \rlap.{' }\else $\rlap{.}' $\fi}
% ' over period.%
\def\arcmper{\ifmmode \rlap.{'' }\else $\rlap{.}'' $\fi}
% " over period.%

\def\aj{{AJ}, }
\def\apj{{ApJ}, }
\def\apjs{{ApJSup}, }
\def\apjl{{ApJLett}, }

\def\mn{{MNRAS}, }

\def\spose#1{\hbox to 0pt{#1\hss}}
\def\lta{\mathrel{\spose{\lower 3pt\hbox{$\mathchar"218$}}
     \raise 2.0pt\hbox{$\mathchar"13C$}}}
\def\gta{\mathrel{\spose{\lower 3pt\hbox{$\mathchar"218$}}
     \raise 2.0pt\hbox{$\mathchar"13E$}}}

\def\hkpc{$h^{-1}$\thinspace kpc}

\font\bb=cmbx12
\ifapj\vglue 29.10pt\else \vglue 2.0truecm \fi
\centerline{\bb  A SURVEY OF FAINT GALAXY PAIRS}
\bigskip
\centerline{R. G. Carlberg$^1$}

\medskip
\centerline{Department of Astronomy, University of Toronto, }
\centerline{Toronto, Ontario, M5S 1A1, Canada}
\medskip
\centerline{C. J. Pritchet$^1$}
\centerline{Department of Physics and Astronomy, University of Victoria}
\centerline{Victoria, BC, V8W 2Y2, Canada}
\medskip\centerline{and}
\centerline{L. Infante}
\centerline{Astrophysics Group, P. Universidad Cat\'olica De Chile}
\centerline{Casilla 104, Santiago 22, Chile}
\footnote{}
{$^1$ Visiting Astronomer, Canada--France--Hawaii Telescope, which is operated
by the National Research Council of Canada, le Centre National de Recherche
Scientifique, and the University of Hawaii.}

\bigskip
\ifapj
\centerline{received: $\underline{\hbox to 6truecm{\hphantom\hfill}}$}
%\vskip 0.5truein
%\centerline{accepted: $\underline{\hbox to 6truecm{\hphantom\null}}$}
\fi

\vfill\eject

\centerline{ABSTRACT}

\bigskip

A sample of faint, V magnitude selected, galaxy pairs, having
physical separations less than approximately 20\hkpc, is used to
examine the rise in the merger rate with redshift and the statistical
relations between close pairs and the field galaxy population.
Redshifts have been obtained for 14 galaxies ($V \le 22.5$) that are
in close ($\theta < 6\arcs$) pairs, along with a comparison sample of
38 field galaxies.  Two color photometry is available for about 1000
galaxies in the same fields.  The average redshift of the $V\le22.5$
field population is 0.36, statistically equal to the average redshift
of 0.42 for the pairs.  The similarity of the two redshift
distributions, $\Delta z\le 0.1$, limits any differential luminosity
enhancement of close pairs to less than half a magnitude.  The pairs
are somewhat bluer than the field and have nearly twice the average
[O~II] detection rate of the field, but the differences are not
statistically significant.  The field population has an angular
correlation at separations of $\theta\le$6\arcs\ higher than the
inward extrapolation of $\omega(\theta)\propto \theta^{-0.8}$, which
may be a population of ``companions'' not present at the current epoch,
or, luminosity enhancement of intrinsically faint galaxies in pairs.
Physical pairs comprise about 7\% of the faint galaxies in our survey
fields. The same physical separation applied to local galaxies finds
only 2.6\% in pairs.  If the rise in close low relatively velocity
pairs with redshift is parameterized as $(1+z)^m$, then $m=2.9\pm0.8$.
If all pairs at low velocities and $r\le 20$\hkpc\ merge, then the
average galaxy mass would be 32\% smaller at $z=0.4$ than locally.

\ls

\ifapj
\bigskip
\noindent
$\underline{\hbox{Subject headings:}}$
				- galaxies: evolution
				- galaxies: formation
				- galaxies: interactions
\fi

\vfill\eject

\ni\ub{1. INTRODUCTION}

\ls
Pairs of galaxies at separations of less than 10--20\hkpc\ comprise a
small fraction, about 3\%, of the present day galaxy population.  Many
close pairs of galaxies (\eg Arp and Madore 1986) are disturbed, sometimes
grossly, by gravitational interaction with a neighboring galaxy
(Toomre 1977, Larson and Tinsley 1978).  The distorted
morphologies can be understood if the infalling galaxies are on orbits
close to parabolic, which, in the presence of dynamical friction, implies
that most close pairs will merge to produce a single galaxy within
$10^9$ years (\eg Toomre 1977, Barnes 1988).  The impact of merging on
galaxy numbers and masses is a small correction, of order 10\%,
if it occurred at a uniform rate in the past; however, a number of
straightforward arguments predict a rapid rise of merging with
redshift within any gravitational instability theory for the formation
of galaxies.  Toomre (1977) shows that if the distribution of marginal
pairwise binding energies is a constant, then the rate of binary
mergers increases into the past as $t^{-5/3}$, corresponding to
$(1+z)^{5/2}$ in an $\Omega=1$ Universe. Somewhat more general
considerations, using the Press-Schechter (1974) formalism and
including a decoupling from the merging hierarchy, indicate that the
rate of merging is proportional to $(1+z)^m$, where $m\simeq
4.5\Omega^{0.42}$ at low redshift (Carlberg 1990) for a fixed mass
spectrum. Extrapolation of present day merger rates using functions
with such a strong redshift dependence can have a dramatic effect on
the galaxy population; such considerations imply that $L_\ast$ galaxies
might be relatively rare beyond $z=1$ (Broadhurst, Ellis and Shanks 1988;
Rocca-Volmerange and Guiderdoni 1990; Carlberg and Charlot 1990; Cowie,
Songaila, and Hu 1991;
Broadhurst, Ellis, and Glazebrook 1992; Carlberg 1992), although
this conclusion is very sensitive to the normalization and redshift
dependence of the merger rate.

Zepf and Koo (1989) undertook a study of faint galaxy pairs
based on KPNO 4m plates. They found 39 paired galaxies with
separations less than 4.5\arcs, drawn from a total population
of 1055 galaxies with $m_J\le
22$.  A comparison with low redshift pair
samples led them to conclude that the merger rate rises as
$(1+z)^{4\pm2.5}$, under the assumption that the pair redshift
distribution is similar to that observed in the field; furthermore,
they noted that many of the pairs showed evidence of distortions that
might arise from tidal effects.  A recent study of close pairs
using HST data (Burkey \et 1994) finds that the merger rate depends
on redshift as $(1+z)^{2.7}$, in agreement with the above.
Indirect evidence that the rate of
merging does indeed rise rapidly with redshift can also be inferred
from a variety of galaxy phenomena believed to be related to mergers
-- for example, quasars (Boyle \et 1988), IRAS galaxies, (Lonsdale \et
1990), and the Butcher-Oemler effect (Dressler and Gunn 1983, Butcher
and Oemler 1984, Lavery and Henry 1988, Lavery, Henry, and McClure 1992).

The redshift distribution of a magnitude-limited survey of pairs of
galaxies could be quite different from that of the field, for at least
three reasons. If paired galaxies are, on average, more luminous than
field galaxies (perhaps as a result of induced star formation)
then their mean redshifts will be higher.
However, a magnitude-limited survey of pairs of galaxies that are
about to merge  may have somewhat lower mean masses than field galaxies (if
pair luminosity enhancement is primarily concentrated in intrinsically low
luminosity galaxies).
On the other hand, if pairs have the same
luminosities as field galaxies, then selecting pairs at a fixed
angular separation favors pairs at low redshift, because
$\xi(\theta d_A(z))$, the real space correlation function, decreases
along the line of sight.  (Here $d_A(z)$ is the angular diameter distance.)
For the limiting case of no luminosity
evolution, the numbers of {\it physical} pairs per unit volume,
$n_p(z)$,  are distributed (for an
$\Omega=1$ Universe) as $$ n_p(z)\propto (1+z)^{-{{3}\over{2}}-\epsilon+\gamma}
[1-(1+z)^{-{{1}\over{2}}}]^{1-\gamma} n^2(z),
\eqno{(1)}
$$ where $n(z)$ is the redshift distribution of single galaxies, the
correlation function is modeled to evolve as $(1+z)^{-3-\epsilon}$
(\eg Efstathiou \et 1991), and the real space correlation function
$\xi\propto r^{-\gamma}$.  For the relatively shallow depths of
interest here, $z\simeq0.3$, the average redshift of the pair
distribution is expected to be about 10\% smaller than that of the
field. We conclude that the average redshift of pairs, compared to
that of the field, may be an interesting constraint on any
differential luminosity evolution of the pairs.  (Nevertheless, the
effect of equation (1) could potentially be a 50\% reduction in mean
redshift for samples that have a significant fraction of their
population spread beyond $z\gta 1$.)

The main goals of this paper are to compare the redshift distribution
of close pairs and the field, and to estimate the rise in the merger
rate with redshift using a CCD selected sample of pairs both with and
without redshifts. The next section discusses our selection procedure
and observations. Section 3 contains our observational results
(redshifts, colors, and angular correlation function).  The
implications of these data for the evolution of galaxies in mass and
luminosity are discussed in
\S4.

\goodbreak
\ls\ls
\ni\ub{2. PAIR SELECTION AND OBSERVATIONS}

\ls

There are several straightforward considerations in constructing a
sample of galaxy pairs for a small redshift survey. The sky density of
physical pairs with separations less than some small angle $\theta$ is
$N_p=N_0^2(f)\int_0^\theta (1+\omega(\theta)) 2\pi \theta d\theta$ at
flux level $f$. The angular correlation function, $\omega(\theta)$, can
be accurately represented as a power law with slope near
$1-\gamma\simeq -0.8$. The amplitude scales with depth in a
approximately Euclidean manner (Maddox \et 1990, Pritchet and Infante
1992) -- \ie as $\omega(\theta)\propto \theta^{-\gamma+1}f^{\gamma/2}$.
The sky density of faint galaxies brighter than magnitude $m$ rises as
$\log{n(<m)} \propto 0.45 m$, or $n_0(>f) \propto f^{-1.125}$.  In the
regime where $\omega(\theta)>1$, the surface density of physical pairs
at fixed angular separation $\theta$ is proportional to
$n^2_0(>f)\omega(\theta)$, or $n_p \propto f^{-1.35}$ -- \ie the number
of pairs brighter than $m$ obeys $\log{N_p}\propto 0.54m$. Thus the
close pair count rises somewhat faster with magnitude than the number
of galaxies -- a fact which has an interesting implication for the
optimal magnitude at which to observe faint physical pairs.  For some
fixed {\it total} observing time $T$ with a multiobject spectrograph,
the optimal exposure time $t$ on each field is clearly that which
maximizes the rate $R = n_p/t$ at which useful pair spectra are
acquired.  Assuming that all spectra above some limiting continuum
signal-to-noise ratio have measurable redshifts, then one finds that, in
the read noise limit, the rate $R$ is given by
$R \propto t^{0.35}$, whereas in the sky
noise limit, $R \propto t^{-0.33}$. Therefore the total number of pairs
will be optimized for pairs selected brighter than the magnitude
at which the spectra begin to be sky-noise dominated ($V \simeq 22$ in our
case).

The field galaxies at $V\simeq22$ have a median redshift of $z\simeq
0.35$ (Colless \et 1990, Lilly \et 1991), sufficient for some redshift
leverage on the merger rate--redshift relation. At this redshift, the
field galaxy population is significantly different than that observed
locally: it is bluer, and the luminosity density is about 3 times
higher (Colless \et 1990, Lilly \et 1991, Eales 1993) (even though the
characteristic luminosity $L^*$ is about the same as observed
locally).

For our purposes
pairs should ideally be defined as those objects that are doomed to
merge; however, for projected data in a continuous clustering
hierarchy the definition of a pair is somewhat arbitrary.  A
reasonable operational definition is to select pair candidates at
angles such that $\omega(\theta)\gta1$ (\ie $\gta50$\% probability
that the pair is physical and not projected).  At $V=22.5$ the
amplitude of the angular correlation function is
$\log{\omega(1^\circ)}\simeq-2.5$, or $\omega(1\arcs)=2$ (Pritchet and
Infante 1992), implying that $\omega(\theta)\ge1$ for
$\theta\le3\arcs$ (for $\omega(\theta)\propto\theta^{-0.8}$).
The pair density on the sky is quite low: the expected number of
paired galaxies with $\theta\le 6\arcs$ (calculated by extrapolating
the Pritchet and Infante 1992 correlation function) is about 1.5 pairs
per $6\times5$ arcminute field available for spectroscopy.  To
increase our chances of obtaining pair spectra, pairs were preselected
from the same plates as used for the ``CFHT North Galactic Pole
Survey'' (Infante and Pritchet 1992).  Because of concerns regarding
the separation of close pairs in the original APM image mode data, we
ran the Kron (1980) galaxy finding and photometry algorithms on scans
of the central regions of the plates. To obtain a reasonable number of
pairs we selected all paired galaxies with separations $\theta \le
8\arcs$ (rather than the 3\arcs\ criterion discussed above). Based
on the Pritchet and Infante (1992) correlation function, one might
anticipate that $\sim {{1}\over{3}}$ of all pairs with separations
of 8\arcs\ and V magnitudes $\le 22.5$ will be physical. For $V \le 21.5$
this fraction will rise to $\sim 60$\%, and to $\sim 80$\% for pairs
chosen with a separation in the range 2--8\arcs.

Chosen pairs were all brighter than $J = 22.5$ in photometry done on the
APM scan data from Infante and Pritchet (1992). We then selected fields to
{\it (i)} maximize the number of pairs (at least 2 pairs per field,
never more than 3), and {\it (ii)} preselect (where possible) pairs
with similar orientation for spectroscopy.
Note that this selection procedure
will bias upwards the number of small separation pairs.
Our original selection procedure (magnitude, separation, and
visual check) identified 110 pairs in an area
equivalent to 80 spectroscopy fields (\ie 1.4 pairs per field).
On the basis of the
average numbers of pairs selected ($\sim$2.5 per field) and expected
($\sim$1.4 per field), our procedure should result in a boost of
approximately 60\% in the number of close pairs over that expected in
purely random fields. {\it However}, we show in \S 3 that
the presence of these preselected pairs in our fields actually
has only a small effect on our computed correlation functions (because
the correlations of close pairs turn out to be much stronger than
expected).

Using the above considerations to choose fields,
observations were acquired at the Canada-France-Hawaii Telescope in
1992 March using the Marlin multi-object imaging spectrograph in
conjunction with a thick, front-side illuminated, Ford Aerospace/SAIC
$1024^2$ CCD (read noise 6.4 $e^-$ pixel$^{-1}$; peak quantum efficiency
$q_{max} \simeq 40$\% at 6500\AA; $q = {{1}\over{2}} q_{max}$ at
5000\AA\ and 8000\AA).
Imaging observations were acquired of seven of our preselected fields,
each approximately $6'\times6'$ in area, through Johnson $V$ and
Kron-Cousins $I$ filters
(exposure times of 15 min in each filter).

The imaging data were used to design a slitlet mask for each field, using
software available at the telescope. First, preselected pairs (see above)
were identified on the images, and as many pair galaxies as
possible were assigned a slit width (1.5\arcs) and
length (nominally 15\arcs).
(It was not possible to obtain spectra of {\it all} galaxies in pairs, since
$\gta$50\% of them had orientations that would have resulted in overlapping
spectra.) Once the pairs had been assigned slits, the rest of each field
was filled with field objects, many of which were below our expected
spectroscopic completeness limit of $V \approx 22$.
The final data file
containing slit coordinates was sent to a computer-controlled laser
punching machine (``LAMA''), which produced the final slitlet mask.
Spectroscopy was then
obtained over a useful wavelength range $\lambda\lambda$5000--8000,
with a dispersion of ~7.7\AA\ pix$^{-1}$
and slits 1.5\arcs\ wide and usually 10\arcs--15\arcs\ long.
The total spectroscopic integration time per field was
2 hours, spread over 4 individual 30 min exposures. Short (60 s)
direct images were taken through the mask (at the beginning
and in the middle of each
2 hour sequence of exposures), to check the alignment of objects
in the masks, and hence test for flexure. Arcs were
acquired at the beginning and end of each 2 hour integration
on a field.

Spectra were extracted using the IRAF\footnote{$^2$} {IRAF is
distributed by the National Optical Astronomical Observatories, which
is operated by the Association of Universities for Research in
Astronomy, Inc., under contract to the National Science Foundation.}
{\sl apextract} package.  Redshifts were obtained by identifying two
or more of [OII] $\lambda$3727, 4000\AA\ break, H$\beta$ $\lambda$4861,
[OIII] $\lambda\lambda$4957,5007, Mg b $\lambda$5180, and in some cases
H$\alpha$ $\lambda$6563. The continuum shape also played a role in
checking redshifts obtained from features and breaks. A quality index
$Q$, ranging from 1 (highest quality) to 6, was assigned to each redshift.
Most of the analysis below refers to redshifts with $Q \le 3$, although
including $Q=4$ redshifts changes none of our conclusions.

The imaging fields were later analyzed using photometry and image
classification loosely based on the Kron (1980) algorithm.
At $V\le22.5$ (and $I\le23$) our survey contains 1062 objects,
of which about 60\% are galaxies, in an
area of about $9\times10^5$ square arcseconds (some area was lost due to
internal reflections from bright stars). Internal photometric errors are
about 0.1 mag at V=22.5 and I=22. The image quality is good, 0.8\arcs, in
the field center, but degrades to 1.5\arcs\ in the field corners.

\goodbreak
\ls\ls
\ni\ub{3. OBSERVATIONAL RESULTS}
\ls

The average properties of the pairs (chosen to
have $\theta \le 6\arcs$)
and field sample are given in
Table~1. The second column gives the number of objects placed in
slits (excluding stars and bad extractions), and the third column
gives the number of objects, $N_z$, for which we were able to obtain redshifts.
The pair redshift sample is 8/14=57\% complete, and the
field is about 28/63=44\% complete (for $V \le 21.5$).
There are at least two reasons for this low completeness.
{\it (i)} The CCD that we used had poor sensitivity in the blue, with
essentially no response below $\sim 4500$\AA, and only $\sim{{1}\over{3}}$ of
peak response at 4800\AA. This made the 4000\AA\ break and [O II] $\lambda3727$
impossible to detect for $z<0.1$ objects, and very difficult at $z\lta0.2$.
%%% LDSS 25/97=26% at z<0.2 (21<J<22.5 --> 20.4<V<21.9)
%%% BES 85/188=45% at z=0.2 (20<J<21.5 -> 19.4<V<20.9)
Interpolating between the results of Colless \et (1990) and Broadhurst
\et (1988), we find that approximately ${{1}\over{3}}$ of the galaxies
in our $V<21.5$ sample should lie at redshifts $z<0.2$, compared with
5/38=13\% actually observed.  Our overall completeness for objects
brighter than $V=21.5$ and with $z \ge 0.2$ therefore rises to $\sim
60$\%. {\it (ii)} The information in Table 1 is a fairly conservative
cut of objects with secure redshifts. Including the next quality level
of redshifts increases the completeness to 61\% at $V \le 21.5$, or
$\sim 80$\% for objects with $z \ge 0.2$. (We note in passing that
such a change in redshift quality has hardly any effect on the mean
redshift, mean color, or mean line fluxes quoted in Table 1.)

%%% MAYBE WE SHOULD ALSO COMPARE OUR
%%% COMPLETENESS WITH THAT OBTAINED BY OTHERS USING THE SAME SETUP!
%%% AFTER ALL, EVEN V=21.5 IS PRETTY FAINT, AND WITH A CRAPPY CCD ...

\ifapj\baselineskip 15pt\fi
\setbox21= \vbox{\parindent 0pt
%\centerline
{Table 1: Average Properties of Paired and Field Galaxies}
\bigskip
\offinterlineskip
{\halign{
\vrule#&\strut\quad#\quad\hfil&\vrule#&
\quad\hfil #\quad&\vrule#&
\quad\hfil #\quad&\vrule#&
\quad\hfil #\quad&\vrule#&
\quad\hfil #\quad&\vrule#&
\quad\hfil #\quad&\vrule#&
\quad\hfil #\quad&\vrule# \cr
\noalign{\hrule}
&    && && && && && && &\cr
&Sample && $N_{slit}$ && $N_z$ && $\langle z \rangle$~~~~~ &&
	$\langle V \rangle~$ && $\langle V-I \rangle$~~ &&
	$\langle \hbox{L(O~II)} \rangle$~~~~~ &\cr
&    && && && && && &&[$h^{-2}$ erg s$^{-1}$]~~~~~ &\cr
\noalign{\hrule}
&    && && && && && && &\cr
%
% my L(O II) numbers seem to be very different from the values
% that were present in the revised version of the table.
% The redshifts changed a little--I put in the values I get
% with the redshifts I currently have. Did the fluxes change as well?
% I think I am using up to date data.
%
% GO WITH YOUR STUFF. I THINK I UNDERSTAND THE DIFFERENCES FOR THE
% <L(OII)>. I AM PUZZLED BY THE DIFFERENCES IN Nz AND Nslit BECAUSE
% I THOUGHT I DID THIS CAREFULLY - BUT IT DOESN'T CHANGE THE RESULTS
% SIGNIFICANTLY. WE WOULD ONLY HAVE TO WORRY ABOUT THIS IN DETAIL IF
% THE REFEREE INSISTS THAT WE INCLUDE A TABLE OF DATA!
%
%& $V\le21.5$ pairs && 14 &&  8 && 0.28$\pm$0.03 && 20.5 &&
%%1.19$\pm0.13$&&$3.8\pm1.3\times10^{40}$  & \cr
%& $V\le21.5$ field && 63 && 28 && 0.36$\pm$0.03 && 20.7 &&
%%1.34$\pm0.09$&&$3.9\pm1.1\times10^{40}$ & \cr
&    && && && && && && &\cr
%& $V\le22.5$ pairs && 24 && 14 && 0.42$\pm$0.05 && 21.1 &&
%%1.41$\pm0.15$&&$7.1\pm1.9\times10^{40}$ & \cr
%& $V\le22.5$ field &&97 && 38 && 0.37$\pm$0.03 && 21.1 &&
%%1.33$\pm0.09$&&$5.2\pm1.4\times10^{40}$ & \cr
& $V\le21.5$ pairs && 14 &&  8 && 0.28$\pm$0.03 && 20.5 &&
1.19$\pm0.13$&&$2.4\pm1.3\times10^{40}$  & \cr
& $V\le21.5$ field && 63 && 28 && 0.36$\pm$0.03 && 20.7 &&
1.34$\pm0.09$&&$1.8\pm1.1\times10^{40}$ & \cr
&    && && && && && && &\cr
& $V\le22.5$ pairs && 24 && 14 && 0.42$\pm$0.05 && 21.1 &&
1.41$\pm0.15$&&$4.6\pm1.9\times10^{40}$ & \cr
& $V\le22.5$ field && 97 && 38 && 0.37$\pm$0.03 && 21.1 &&
1.33$\pm0.09$&&$2.3\pm1.4\times10^{40}$ & \cr
\noalign{\hrule}
\noalign{\medskip}
\multispan9{Note: all errors are errors in the mean.\hfil}\cr
}
}}

\ifapj\baselineskip 24pt\fi
\ifapj\else \midinsert \medskip \copy21 \medskip \endinsert \fi

The fourth column of Table~1 gives the average redshifts for $V <21.5$
and $V < 22.5$; these are a little higher than the values of 0.25 to
0.35 expected for the two magnitude ranges (Colless \et 1990).  Apart
from one extragalactic HII region, the lowest redshift detected is
0.137, a consequence of the poor sensitivity of our CCD below 4500\AA,
as discussed above. The dispersion about the mean redshift is about
0.16 for both field and pair samples.

\medskip\goodbreak
\centerline{3.1 \it Redshift Distribution}
\smallskip

The most significant result arising from the redshift data is that
{\it the average redshifts of the pair and field samples are the same
to within $\Delta z/z\lta0.1$}.  This requires that both populations
have similar characteristic luminosities: the intrinsic luminosity of
the two samples is identical to within 40\%. (Formally the $V < 21.5$
sample places an even more stringent constraint on luminosity
evolution of the pairs sample, because $\langle z \rangle_{pair} <
\langle z \rangle_{field}$. Nevertheless, this result must be viewed
with some caution because of the small number of redshifts for
$V<21.5$ pairs, and because of subclustering in
these redshifts.)  Overall, our redshifts place a powerful constraint
on differential luminosity evolution between the two populations. (As
noted in \S 1, the bias introduced into pair selection because of the
correlation function is much smaller than this.)

\medskip\goodbreak
\centerline{3.2 \it Colors and Luminosities}
\smallskip

Column 5 of Table 1 gives the average magnitude of the galaxies, and
column 6 gives the mean color and the standard deviation of the mean.
In the $V\le21.5$ sample the pairs are bluer than the field, but the
difference is at the $1\sigma$ level. (The colors have not been $k$-corrected,
but this should have little effect on the above result because the mean
redshifts of the samples are the same.)
Color differences of $\vert\Delta(V-I)\vert \gta 0.2$
magnitudes are nevertheless ruled out.

The average luminosity in the
[O II] 3727\AA\ line, column 7, is calculated (assuming that
non-detections have zero line luminosity) in an $H_0=100h, \Omega=1$
cosmology. There is no significant difference in this quantity for the
field and pair samples; this result is perhaps not too surprising, given
the fact that fewer than 40\% of the pairs and 20\% of the field
galaxies have detections. Perhaps the only difference that is notable
in the [O II] $\lambda3727$ properties of the survey is that detections
are twice as frequent in the pair sample as in the field sample
(5/14=36\% for pairs versus 13/70=19\% for field objects at $V<21.5$;
9/24=38\% for pairs versus 16/104=15\% for field objects at $V<22.5$).

There are 7 pairs for which both members of the pair possess redshifts;
5 of these pairs belong to our ``pre-selected'' sample, whereas 2 were
``serendipitous''. (One ``pair'' is actually a triple.)
(Since our sample of pairs was selected using the [very faint] cutoff
magnitudes $V\le22.5$ and $I\le23$, and because of orientation effects,
it is not surprising that many pairs possess only a single redshift.) Three
of the pairs with full redshift information have $\Delta z \le 0.002$, one
pair has $\Delta z=0.031$, and the other two are separated by $\Delta
z>0.2$:  that is, three pairs are physical, whereas four are projected.
The redshifts of the physical pairs are 0.378, 0.412, and 0.586.
Considering
that pairs were selected on the basis that $\omega(\theta)\gta 0.5$, it
is entirely to be expected that of order half of the pairs
should be physically associated.

\ifapj\baselineskip 15pt\fi
\setbox22= \vbox{\parindent 0pt
%\centerline
{Table 2: Angular Correlations}
\bigskip
\offinterlineskip
{\halign{
\vrule#&\strut\quad#\quad\hfil&\vrule#&
\quad\hfil #\quad&\vrule#&
\quad\hfil #\quad&\vrule#&
\quad\hfil #\quad&\vrule# \cr
\noalign{\hrule}
&       &&   && 	       && &\cr
&	&& $V_f<21.5$ &&$V_f<22.5$ &&$ I_f<21.5$ &\cr
%subscript f is for r_2>1.8, other is r_2>1.0
% 0.37 arcsec/pixel
%%% I HAVE INTERCHANGED THE V<22.5 COLUMN AND I<21.5 COLUMN - SEE TEXT
&       &&   && 		&& &\cr
\noalign{\hrule}
& $\omega(<6\arcs)$  && $4.80\pm1.19$ &&$2.06\pm0.41$  &&$1.07\pm0.29$& \cr
& $\omega( 6\arcs-24\arcs)$ && $0.25\pm0.14$ && $0.09\pm0.06$ && $0.18\pm0.05$&
\cr
&       &&   &&	&& &\cr
& $\omega(<6\arcs)/\omega(6-24\arcs)$ && 19.2 && 24.0&& 5.9 & \cr
&       &&   &&	&& &\cr
& $N_p(<6\arcs)$&& 24 && 56 &&52 & \cr
& $N_p(6\arcs-24\arcs)$&& 80 && 306 &&514& \cr
&       &&   &&	&& &\cr
& $N_p(<6\arcs)/N_p(6-24\arcs)$ && 0.30$\pm$0.07  && 0.18$\pm$0.03 &&
0.10$\pm$0.02 & \cr
\noalign{\hrule}
}
}}

\ifapj\baselineskip 24pt\fi
\ifapj\else \midinsert \medskip \copy22 \medskip \endinsert \fi

\medskip\goodbreak
\centerline{3.3 \it Angular Correlations}
\smallskip

Table~2 gives the data for the small separation end of the angular
correlation function, derived from our imaging observations.
The samples marked with a subscript $f$ have
been selected to be ``fuzzy'': that is, their Kron (1980) $r_{-2}$
parameters have been preselected (using $r_{-2}\ge1.8$ for this data)
to remove (unclustered) stars. (Note that there may be some compact
galaxies which are excluded by this criterion.)

\setbox11=\vbox{
\ifapj\else\baselineskip 10pt\fi %\eightrm\fi
\noindent
Figure 1:\hskip 5mm $V$ (left member) and $I$ (right member)
images of 37\arcs\ regions around $\theta\le$6\arcs\ pairs.}

\ifpairpics
\setbox31=\vbox{
\hbox{
\epsfysize 1.85truecm
\epsfbox[46 658 536 776]{pics/2J2_1.sp}
\hfil
\epsfysize 1.85truecm
\epsfbox[45 658 536 776]{pics/2J2_2.sp}}
\vskip 1.74truecm
\hbox{
\epsfysize 1.85truecm
\epsfbox[46 658 536 776]{pics/2J2_3.sp} %star
\hfil
\epsfysize 1.85truecm
\epsfbox[45 658 536 776]{pics/2J5_1.sp}}
\vskip 1.74truecm
\hbox{
\epsfysize 1.85truecm
\epsfbox[46 658 536 776]{pics/2J6_1.sp}
\hfil
\epsfysize 1.85truecm
\epsfbox[45 658 536 776]{pics/2J6_2.sp}}
\vskip 1.74truecm
\hbox{
\epsfysize 1.85truecm
\epsfbox[46 658 536 776]{pics/2J6_3.sp}
\hfil
\epsfysize 1.85truecm
\epsfbox[45 658 536 776]{pics/2J6_4.sp}}
\vskip 1.74truecm
\hbox{
\epsfysize 1.85truecm
\epsfbox[46 658 536 776]{pics/3J1_1.sp}
\hfil
\epsfysize 1.85truecm
\epsfbox[45 658 536 776]{pics/3J1_2.sp}}
\vskip 1.74truecm
\hbox{
\epsfysize 1.85truecm
\epsfbox[46 658 536 776]{pics/3J1_3.sp}
\hfil
\epsfysize 1.85truecm
\epsfbox[45 658 536 776]{pics/3J5_1.sp}}
}

\setbox32=\vbox{
\hbox{
\epsfysize 1.85truecm
\epsfbox[46 658 536 776]{pics/3J5_2.sp}
\hfil
\epsfysize 1.85truecm
\epsfbox[45 658 536 776]{pics/3J5_3.sp}}
\vskip 1.74truecm
\hbox{
\epsfysize 1.85truecm
\epsfbox[46 658 536 776]{pics/3J5_4.sp}
\hfil
\epsfysize 1.85truecm
\epsfbox[45 658 536 776]{pics/3J5_5.sp}}
\vskip 1.74truecm
\hbox{
\epsfysize 1.85truecm
\epsfbox[46 658 536 776]{pics/4J7_1.sp}
\hfil
\epsfysize 1.85truecm
\epsfbox[45 658 536 776]{pics/4J7_2.sp}}
\vskip 1.74truecm
\hbox{
\epsfysize 1.85truecm
\epsfbox[46 658 536 776]{pics/4J7_3.sp}
\hfil
\epsfysize 1.85truecm
\epsfbox[45 658 536 776]{pics/4J7_4.sp}}
\vskip 1.74truecm
\hbox{
\epsfysize 1.85truecm
\epsfbox[46 658 536 776]{pics/5J6_1.sp}
\hfil
\epsfysize 1.85truecm
\epsfbox[45 658 536 776]{pics/5J6_2.sp}}
\vskip 1.74truecm
\hbox{
\epsfysize 1.85truecm
\epsfbox[46 658 536 776]{pics/5J6_3.sp}
\hfil
\epsfysize 1.85truecm
\epsfbox[45 658 536 776]{pics/5J6_4.sp} %dummy for plot symmetry
}
}

\ifapj\else
\pageinsert\copy31
\vfill\copy11\endinsert \fi

\ifapj\else
\pageinsert\copy32
\vfill\copy11\endinsert\fi
\else\fi

A detailed calculation shows that the $\theta<6\arcs$ bin should have
about $15\times$ fewer pairs than the larger $6<\theta<24\arcs$ bin
(assuming a power-law angular correlation function with power-law slope
$\delta=0.8$ and $\omega(3\arcs)=1$, and that galaxies closer than
$3\arcs$ cannot be distinguished).  (Note that we find no pairs with
separations $<1.5\arcs$.)  The observed ratios of pairs are in fact
$\sim$3--5 times larger than this for the $V$ samples (with the larger
excess arising from the most reliable $V_f < 21.5$ sample). The
correlation function, $\omega(<6\arcs)$, is expected to be 2.6 times
larger than $\omega(6\arcs-24\arcs)$ (under the same assumptions as
above).  In fact the observed correlations in the $V$ band are a factor
of 7--9 times larger than expected in the innermost bin (a factor of 7
for the $V < 21.5$ sample). Excess correlations for the $I_f<21.5$
sample are smaller, but still statistically significant (approximately
a 2$\sigma$ excess in the pair ratio $N_p(<6\arcs)/N_p(6-24\arcs)$).
The lower excess correlations in the $I$ band could be due to the
higher mean redshift of this sample, or to the fact that the pairs
are slightly bluer than the field (at least for the $V<21.5$ sample).

These excesses of close pairs (over what would be expected from an
extrapolation of $\omega(\theta)$ at larger $\theta$) are even more
significant if $\delta < 0.8$ (\eg Maddox \et 1990), or if (as might be
expected) there is incompleteness in the $\theta < 6\arcs$ bin.  (At
the mean redshift of the sample, 6$\arcs$ corresponds to a physical
separation of 19\hkpc\ for $\Omega=1$.) It is interesting to note that
there is a weak indication of excess small scale correlation at similar
separations at {\it low} redshifts (Davis and Peebles 1983), although
the statistical significance of this result is unclear.

The bias that would result from our field selection (\S 2) appears to
fall {\it far} short of explaining the observed excess of close pairs.
Only two of our preselected pairs survive the separation ($\Delta\theta
< 6\arcs$) and morphology ($r_{-2}>1.8$) cuts that were used in
Table~2; removing these two pairs would have no effect on the
significance of the excess of correlation power at small separations.

The images of all the close pairs
%(Figure~1)
were examined for signs
of distortions and interactions.  Although some of the galaxies have
features that could be created in tidal distortions, few of the
galaxies are strongly distorted. A sample of {\it nearby}
pairs has been constructed from the UGC catalog (Nilson 1973)
with the same projected physical separation;
only a small percentage of the equivalent close pairs
of galaxies are classified as peculiar (3 of 140) or uncertain (6
of 140). (This nearby pair sample
is discussed further below.)

\goodbreak
\ls\ls
\ni\ub{4. IMPLICATIONS FOR MERGER RATES AND MASS EVOLUTION}
\ls

In this section we examine the consequences of our observations
for simple merger models of galaxy evolution. We argue below that
mergers {\it must} be taking place in our pair population, and from
this infer that mergers must have taken place at a significant rate
in the field population as well. Formally, none of our observations
(mean redshift, colors, line strengths) {\it demand} that these mergers
cause star formation. However, we argue below that our observations
are in fact consistent with models in which mergers induce evolutionary
brightening of both the paired galaxies and the field population.

\medskip\goodbreak
\centerline {\rm 4.1 \it Pair Fraction and Merger Rate}
\smallskip

The fraction of galaxies that are members of $\theta\le 6\arcs$ pairs
is about 14\% of our sample (30 of 196 for $V<21.5$, and
52 of 410 for $V<22.5$).  We estimate that about 1/2 of these are
physical pairs, based on our redshift data (3 of 7 physical pairs) and the
measurement that $\omega(\theta)\simeq 2$ for these pairs.  Therefore
about 7\% of the faint galaxies in our sample at $z\simeq 0.4$ are
in physical pairs with separations projecting to 19\hkpc.

%%% RAY, CHECK THE FOLLOWING: RELEVANT DATA IS THAT BURKEY ET AL FIND
%%% 34% OF THE GALAXIES WITH 18<I<22 BELONG TO PAIRS WITH SEPARATIONS
%%% 0.5"-4.0"
%looks ok to me
How does this result compare with the fraction of close pairs found
by Burkey \et (1994) in deep HST images? Their fraction of galaxies
in pairs is 34\%, for galaxies in the magnitude range $18<I<22$
and separations $0.5\arcs-4\arcs$. If we assume $\omega(\theta)\simeq2-5$
for $\theta \le 6\arcs$ (as in Table 2), and extrapolate $\omega(\theta)$
to smaller separations using $\delta=0.8$, we would predict fractions of
galaxies in pairs around 12--14\% -- a factor of 2.6$\times$ smaller than
observed by Burkey \et. Therefore the fraction of galaxies in close pairs,
and the correlation function, is increasing much more steeply at small
separations than a canonical power-law slope of $\delta=0.8$ would predict.
This is broadly consistent with the fact that our data also shows excess
correlation power compared to a power-law fitted at larger separations.

At
separations of $\lta20$\hkpc, interacting pairs should quickly merge. Toomre
(1977) estimates that the time to merge is typically 0.5 Gyr for such
objects; this estimate should be relatively conservative, since the
orbital time for a 200 \kms rotation velocity is 0.4$h^{-1}$ Gyr at
this distance.  The merger timescale for the population of faint
galaxies as a whole is therefore $t_{mg}(z=0.4) \simeq 0.5 f_{mg}^{-1}$ Gyr, or
7.1 Gyr. For $H_0=50$, this is essentially identical to the Hubble
time at $z=0.4$ (7.9 Gyr), implying that such a large rate of merging
will have a significant impact on the average masses of galaxies. Note
that all merger rates scale with the assumed 0.5 Gyr for pairwise
mergers. This is likely to be a lower limit, with mergers occurring
more slowly in groups where tidal fields help to delay merging.

A low redshift sample with comparable pair properties to our faint
galaxy sample can be constructed from the (angular diameter limited)
UGC catalog, which is statistically complete to $B=14.5$ (Nilson
1973).  The average redshift (of the galaxies in the Nilson catalog
which have redshifts) is 0.007 (2100 km s$^{-1}$). At this redshift 19\hkpc\
projects to 192\arcs.  Restricting the catalog to objects with
$B<14.5$, with measured sizes in both $B$ and $R$, we find 140 paired
galaxies of 3066 in the sample, or a fraction of 4.6\%. Taking all of
these to be physical pairs, we estimate $t_{mg}(z=0)=11$ Gyr, where we
again assume that the timescale for an individual merger is 0.5 Gyr (as for
the faint pair sample).

Although projection is not a problem for nearby galaxy pairs, some
physically close galaxies will be moving at such high relative
velocities that they are unlikely to merge.  Of the 70 UGC pairs,
there are 23 for which both galaxies have redshifts; only 13 of these
have line of sight velocity differences less than 350 \kms\ and hence
are likely to merge (the average velocity difference of the
``$<350$ \kms'' group is 107 \kms, compared to 570 \kms\ for the ``$>350$
\kms''
group).  Applying this fraction, 0.56, to the 140 close galaxies,
reduces the number likely to merge to 79, which decreases $f_{mg}$ to
2.6\%, and increases the merger timescale to 19 Gyr. It is interesting
to note that 57\% of the UGC close pairs are E or S0 galaxies, whereas
only 16\% of the catalogue has these morphological types.  Not
surprisingly, 80\% of the high velocity pairs are E or S0 galaxies.  It
is noteworthy that none of the galaxies in the 13 close pairs with small
pairwise velocities is
classified peculiar, although there are 2 irregulars.
%awk '{d=$8-$9;if(d<0)d=-d;if(d<=350) print $0}' tmp
%   30    1181  1182    5628  5631   E         Sb          1338   1111
%   33    1270  1271    5943  5947   S(c)      IRR         1656   1645
%   34    1296  1297    6018  6019   X         IRR          965   1014
%   39    1594  1598    6949  6957   S?        S           3133   3362
%   47    1972  1973    7793  7794   Sc        Sc          2278   2253
%   49    2021  2022    7913  7915   Sc        E           1396   1269
%   52    2219  2220    8659  8663   S         S           6810   6635
%   55    2361  2362    9160  9161   Sa        Sb-c        3165   3175
%   56    2373  2374    9199  9201   S0?       E           1716   1528
%   58    2507  2509    9742  9746   Sa        SBb         4705   4664
%   59    2540  2541    9921  9922   S0        S(c)        2100   2140
%   62    2665  2667   10742 10745   S         S0-a        3064   3283
%   67    3042  3043   12832 12834   X         SBa         4338   4282

The timescale for merging at the current epoch found here, 19 Gyr, is
substantially shorter than implied in Toomre's (1977) discussion (but
see Carlberg and Couchman 1989).  The principal difference between
these two results is that Toomre demanded that merging objects be
severely distorted, with long tidal tails. In fact, there are many
more interacting objects in the Arp and Madore (1986) catalog (by a
factor of 5 or so) which are likely to merge, yet which lack such
conspicuous evidence of interactions.

\medskip
\centerline {\rm 4.2 \it Evolutionary Effects}
\smallskip

%Larson&Tinsley
% Age 	M/L (const) B-V		M/L(V) (burst)	B-V	M/L(B)
% 0.1	0.05	    0.04	0.11		0.26	0.14
% 1.0	0.20	    0.27	0.69		0.51	1.10
% 10.	0.97	    0.50	3.8		0.94	9.0
% 20.	1.7	    0.56	8.1		1.02	20.7
% the burst M/L is roughly fit by: M/L= 0.7 (t/Gyr)^0.75
% L_b/<L_b> (t_i/t_f=T<1) = 0.25 T^-0.75 (1-T)/(1-T^0.25)
% awk '{T=$1;T4=sqrt(sqrt(T)); print 0.25*T4*(1./T-1.)/(1.-T4) }' data
% for T=0.08 get 3.26, T=0.125 get 2.57
% in the blue, M/L is fit by  1.1 (t/Gyr)^0.9
% L_b/<L_b>(blue) = 0.1 T^{-0.9} (1-T)/(1-T^{0.1}), where T=t_i/t_f

The similarities of the faint galaxy pair and field populations have two
possible explanations. Either the presence of a close companion
has no effect on the color or luminosity of a galaxy, or, if
a close companion leads to interactions which boost the
star formation rate (as argued by Larson and Tinsley 1978), then
interactions must be so frequent that
most faint galaxies in the field have recently undergone
interactions.
At redshift $z\simeq0.3$ the luminosity density, roughly
proportional to the star formation rate, is about 3 times higher than at
the current epoch (Broadhurst \et 1988, Lilly \et 1991, Eales 1993).
The following crude calculation shows that the
small fraction of galaxies in close pairs, $\sim$7\% here, could be the
source of all the enhanced star formation in the field.  The Larson
and Tinsley (1978) model for the B luminosity of a burst is
approximately fit by $$ L_b/M = 0.9 t^{-0.9},
\eqno{(2)}
$$ where t is measured in units of Gyr, $M/L$ is in solar units,
and $L$ is measured in the (rest-frame) $B$ band.  The ratio of the total
luminosity of a galaxy having a star formation burst of age
$t_b$ to the average luminosity of post-starburst galaxies of maximum
age $t_m$ is
$$
{L_b\over{\langle L\rangle}} ={{T^{-0.9}+M_0/M_b}\over
10{{1-T^{0.1}}\over{1-T}}+M_0/M_b},
\eqno{(3)}
$$
where $T=t_b/t_m$, and $M_0/M_b$ is the ratio of the mass of the underlying
pre-existing population to the
mass of stars formed in the burst.

Equation~(3) allows an upper limit to be set on what fraction of
an average pairs's stellar mass is being formed in the interaction,
based on
keeping $L_b/\langle L\rangle\le 1.58$ (0\mper5, the limit implied by the
redshift distributions).  For interactions occurring at a uniform rate,
$t_b/t_m$ can be set to the fraction of interacting
galaxies, 7\%. Using Equation~(3) with
$t_b/t_m=0.07$ gives $M_0/M_b\ge12$, which
sets the upper limit to the fractional burst mass,
$M_b/M_0$, as less than 8\%.
If the underlying old population continually forms stars,
then it has  a higher luminosity per unit mass than
a old pure burst, which could easily lower this limit by a factor of 2.
Therefore, induced star formation in close pairs cannot
dramatically change the mass of a galaxy--close pairs at $z\simeq0.4$
are not ``protogalaxies''.

The cumulative effect of interactions over the lifetime of the galaxy
could be a dominant source of its stellar mass, if each interaction
does induce the allowed limit of star formation. That is, if 8\%  (or
4\% for continuous star formation) of
the stellar mass is formed in each interaction, and all the field
galaxies are ``post interaction'', which is allowed since the ratio of
the interaction time of 0.5Gyr to the age of the universe at $z\simeq
0.4$ is about 0.06, then the upper limit to the accumulated burst mass
is 100 (50)\%. Therefore, most of the mass of a typical galaxy could be
built up as a result of interaction induced star formation (assuming
that sufficient gas is always available).  Greatly improved
constraints will be possible with a larger sample.  The data presented
here includes the ``interaction formation'' hypothesis only as an
upper limit.

\medskip
\centerline {\rm 4.3 \it Redshift Dependence of Merger Rate}
\smallskip

The {\it relative} rates of merging at $z=0$ and $z=0.4$ can be
compared without knowing the timescale for merging. We estimate that
7.0$\pm$1.5\% of the galaxies at $\langle z \rangle \simeq 0.4$ are in
pairs with the same physical separation and merger probability, and
2.6$\pm$0.3\% of the galaxies at $z\simeq0$.  The ratio of merger rates
at these two redshifts is therefore $2.7\pm0.7$, where the
uncertainties have been computed by assigning $N^{1/2}$ errors to the
number of objects at each redshift. (Note that the error in the
fraction of objects in pairs at z=0.4 may be larger than derived from
$N^{1/2}$ errors, because of uncertainties in the fraction of pairs
that are physical.) Assuming a $(1+z)^m$ increase in merger rate with
redshift therefore implies that $m=2.95\pm0.8$, in good agreement
with $m=2.7$ derived by Burkey \et (1994) from HST images.

The merger rate index can be tentatively linked to $\Omega$; low
$\Omega$ gives very little growth of structure at late times, although
$m>2$ for any reasonable $\Omega$. The key assumption is that galaxies
moving at less than some critical pairwise velocity will
merge (Aarseth and Fall 1980); the critical pairwise velocity
is comparable to the escape speed. Under the assumption
that the pairwise velocity of encounter of two galaxies evolves as
expected in hierarchical clustering, $(1+z)^m$, and that the critical
velocity remains constant (not completely true for an evolving mass
distribution) then it is straightforward to show that the merger rate
index, $m$, should vary as $m\simeq4.5\Omega^{0.42}$ (Carlberg 1990)
for an unevolving mass spectrum. The rate of change will be reduced if
the galaxies become significantly lighter in the past. Allowing for a 32\%
mass difference between redshift 0 and 0.4 (see below) changes the relation
to $m=3.2\Omega^{0.57}$.  For our measured $m$, the implied
$\Omega=0.87\pm0.4$.  The large random error means that this is not a
very useful constraint on $\Omega$; clearly a larger sample of pairs could
rectify this problem.  The other merger model parameters have relatively little
effect over this range of redshifts.  Certainly the rapid rate of
change in merger rate found here argues against a low $\Omega$
Universe; in such Universes the growth of perturbations ``freezes
out'' at high redshift ($z \approx \Omega^{-1}$), so that the present
day merger rate is slower and depends more weakly on $z$.

\setbox12=\vbox{\narrower
\ifapj\else\baselineskip 11pt\fi %\eightrm\fi
\noindent
%Figure 2:\hskip 5mm The merger rate -- redshift relation. The
Figure 1:\hskip 5mm The merger rate -- redshift relation. The
$z=0.007$ data comes from close pairs in the UGC catalog. The $z=0.4$
results are from this survey. The lines use the theory discussed in
the text, which is based on a generalization of the Press-Schechter
formula.  The size of the ellipses indicates the approximate spread in
the data along each axis. The theoretical predictions ({\it solid
lines}) depend primarily on $t_{mg}$, which is assigned values of 13, 26,
and 45 Gyr (top to bottom).  The two closely-separated lines in the
middle show the result of changing $v_{mg}$ from 280 to 140 \kms.
Integration under either of the middle lines implies that the average
galaxy at $z=0.4$ has a mass that is 68\% of that of present-day
galaxies. Extrapolating to $z=1$ reduces galaxy masses to 36\% of
present-day values, {\it assuming that low relative velocity pairs merge.}}

\ifpics
 \ifapj\else
 \midinsert
 \epsfysize=12truecm
 \centerline{\epsfbox[48 206 568 678]{mg.ps}}
 \medskip\copy12\endinsert
 \fi
\else\fi

The merger rates estimated from the UGC catalog and our survey are
displayed in Figure~1, along with several parameterizations of the
theoretical merger rate-redshift relation (Carlberg 1990). The merger
rate per galaxy is given as $$ R_{mg} = {t_0\over { c_n t_{mg}}}
\gamma{\left(1.5,0.5 {v_{mg}^2 \over{v_p(z)^2}}\right)}{1 \over t},
\eqno{(4)} $$ where $\gamma(a,x)$ is the incomplete gamma function (the
integral of a Maxwellian speed distribution), $v_{mg}$ is the critical
velocity for merging, taken to be 280 \kms, and $v_p$ is the pairwise
velocity dispersion of galaxies, taken to be
$600(1+z)^{(n-1)/(2n+6)}$\kms (Carlberg 1990). The constant $c_n$
normalizes the expression to $t_{mg}^{-1}$ at $t=t_0$.  We take
$n=-1.5$ as is indicated by the correlation function and the CDM
spectrum for mass scales of 3--10 galaxy masses.  This value of $n$
predicts that the pairwise velocity dispersion evolves as
$v_p(z)\propto (1+z)^{-5/6}$, a 25\% drop out to redshift 0.4. The
critical velocity for merging will evolve as the galaxy masses change;
however for $v_{mg}\propto M^{1/3}$ the inferred drop of $\sim30$\% in
the average galaxy mass (see below) is sufficiently small that changes
in $v_{mg}/v_p$ are caused primarily by changes in $v_p$.  For small
values of the $t_{mg}$ this expression exceeds the rate of halo
mergers, $$ R_{mg} = {2 \over {n+3}} {1 \over t}, \eqno{(5)} $$
predicted at larger $z$. The $R_{mg}$ plotted in Figure~1 is the
minimum of the two expressions, with Equation~(4 )dominating in the
$0\le z\le 0.4$ range where our data lies. The scaling with redshift of
Equation~(4) should be about right, but the normalization is subject to a
lack of almost any knowledge of the details of merging of galaxy size
objects in a cosmological context.

The change in the numbers of galaxies as a result of merging is $$ {d
n_g \over{ d t}} = R_{mg} n_g
\eqno{(6)}
$$ Under the assumption that the baryonic mass content of galaxies may
be rearranged but not changed, Eq. 4 implies that the characteristic
mass of galaxies decreases as $$ M_\ast(z) = M_\ast(0)
\exp{[-\int^{t(z)}_0 R_{mg} \, dt]}
\eqno{(7)}
$$ with redshift. We make a smooth transition between the two rates by
using an inverse quartic interpolation.  The model line shown in
Figure~1 implies that at $z=0.4$, the characteristic mass of galaxies
$M_\ast(z=0.4)= 0.69M_\ast(0)$. Extrapolating to $z=1$ predicts that
$M_\ast(z=1)=0.36M_\ast(0)$, which is indeed a dramatic change in the
population. (However, it should be noted that the uncertainty in this
result is very large -- nearly a factor of 2 -- and depends
on the accuracy of the theoretical model.) These estimates of mass changes can
be
taken as upper limits, since the time to merge may well be
longer than the 0.5 Gyr assumed, and since some close pairs may fail to
merge, as a result of strong gravitational tides in clusters and
groups.

\medskip
\centerline {4.4 \it Companion Galaxies?}
\smallskip

The excess small scale correlations discussed in \S 3 (about
a factor of 2 or 3 more galaxies at $\theta\le6\arcs$ than
expected, a result which is more prominent in V than I) have at least
two possible explanations consistent with our observed redshift
distribution: either they represent a true density excess, or
they are due to a selective
brightening only of intrinsically low luminosity companions.
At low redshift the correlation function appears to be
consistent with a power law on all scales (Lake and Tremaine 1980,
Davis and Peebles 1983). At somewhat higher redshifts
quasar absorption line studies suggest that galaxies may
have associated substructure (Wolfe \et 1986, Steidel 1993).
A luminosity excess for close pairs is
{\it predicted} in models in which the changes in the field population
luminosity density at $z\simeq0.3$ are due to galaxy interactions and
merging (Rocca-Volmerange and Guiderdoni 1990, Cowie \et 1991,
Broadhurst \et 1992, Carlberg and Charlot 1992, Carlberg 1992).  For an
unperturbed luminosity function of the form $L^{-\alpha} d\ln{L}$,
with $\alpha\simeq1$ (approximately the normal Schechter function
value) a brightening of a factor of 2, as allowed by our limits on
the redshift distributions, increases the numbers above some limiting
$L$ by at least a factor of 2.

Our data is consistent with either a density excess or a companion
luminosity enhancement.  A large, photometrically accurate imaging and
spectroscopic survey will be necessary to accurately define the mean
properties of these populations, which have large intrinsic
dispersions.  At the optimal magnitude for pair surveys, $V\simeq
21-22$, there are only 100 pairs per square degree, requiring a great
deal of sky to be surveyed.

\ls\ls\goodbreak\ni\ub{5. CONCLUSIONS}

\ls
We have 53 redshifts of galaxies drawn from a $V\le22.5$ photometric
sample of $\simeq1000$ galaxies.  Close pairs are defined as objects
closer than 6\arcs\ on the sky, for which we have 14 redshifts.  The
pairs are compared to a sample of nearby pairs assembled from the UGC
catalog with the same projected physical separation.  The principal
result of this study is that the mean redshift of the pair population is
very similar to the mean redshift of the field galaxies at the same
magnitude limit. This limits the differential luminosity evolution between
the field and pairs to less than $\sim 0.5$ mag difference. The small
magnitude difference implies that if  an average pair has
interaction induced star formation, it only adds 8\% more stellar mass
on the average. Integrated over the lifetime of the galaxy this could
account for the entire stellar mass, but as an upper limit, given
the observational constraints we have derived.

A second result is that the rise of the fraction of galaxies in close
physical pairs implies that the merger rate--redshift relation is
$(1+z)^{2.9\pm0.8}$ over the interval $0<z<0.4$. This result is
somewhat shallower than that of Zepf and Koo (1989); however it should
be more reliable because uncertainties concerning the redshift
distribution of the pairs have been removed. When combined with
theoretical modeling of the expected merger rate-redshift relation,
our observations have two important implications.  If all of these
galaxies do merge, then the average mass at $z\simeq0.4$ is reduced by
32\%, and by a factor of $\sim2.7$ at redshift 1 (the latter result
depending significantly on the accuracy of the model used for the
extrapolation). Secondly, the slope of the merger-rate redshift
relation indicates $\Omega\simeq0.87\pm0.4$. The combination of
model dependencies and large random error makes this density
estimate less than compelling.

There is an excess correlation at projected separations of
$<20$\hkpc. The strong correlation
at $r\lta20$\hkpc\ could be either a local bound population,
or the result of a luminosity enhancement of intrinsically
faint galaxies in  close pairs,

There are several limitations inherent in this study. A larger photometric and
redshift sample would allow a conclusive check of the color
differences between field and pairs, which are expected to be in the
range of 10\%, because the field is so much bluer than observed locally. In
this study there is no conclusive direct indicator that mergers have taken
place, or will take place, other than the (difficult to quantify) distortion
morphology; nor is there any direct indication that pairs have excess
star formation rates.  The relative rate of merging is established to
within an accuracy of a factor of two, but the absolute rate is still
poorly known, and the impact of merging on the mass function hinges
upon an integral over the absolute rate.

\ls\ls\goodbreak
\ni\ub{ACKNOWLEDGEMENTS}

We are grateful to John Hamilton for assistance at the telescope,
to O. LeF\`evre
for assistance with the Marlin instrument, and to Erica
Ellingson and Howard Yee
for interesting and helpful discussions. This work was
supported in part by operating grants from the Natural
Sciences and Engineering Research Council of Canada,
and by FONDECYT project 1930570.

\goodbreak\ls\ls
\ni\ub{REFERENCES}

\ls
%\hi{Arp}
\hi{Aarseth, S. J. \& Fall, S. M. 1980, \apj 236, 43}
\hi{Arp, H. A. \& Madore, B. F. 1986, {\it A Catalogue of Southern
	Peculiar Galaxies and Associations} (Cambridge: Cambridge University
	Press)}
\hi{Barnes, J. E. 1988, \apj 331, 699}
%\hi{Bean, A. J., Efstathiou, G., Ellis, R. S., Peterson, B. A. \& Shanks,
%	T. 1983, \mn {205}, 605}
\hi{Broadhurst, T. J., Ellis, R. S. \& Shanks, T. 1988, \mn {235}, 827}
\hi{Broadhurst, T. J., Ellis, R. S. \& Glazebrook, K. 1992,
	Nature 355, 55}
%%%\hi{Bruzual, G. \& Charlot, S. 1993, \apj in press}
\hi{Boyle, B. J., Shanks, T., \& Peterson, B. A. 1988, \mn
        235, 935}
\hi{Burkey, J. M., Keel, W. C., Windhorst, R. A., \& Franklin, B. E.
	1994, ApJL, submitted}
\hi{Butcher, H., \& Oemler, A. 1984, \apj 285, 426}
%\bibitem{Clam}Carlberg, R. G. 1991, \apj {\bf 375}, 429
\hi{Carlberg, R. G. \& Couchman, H. M. P. 1989, \apj {340}, 47}
\hi{Carlberg, R. G. 1990, \apjl {359}, L1}
\hi{Carlberg, R. G. 1992, \apjl {399}, L31}
\hi{Carlberg, R. G. \& Charlot, S. 1992, \apj {397}, 5}
\hi{Colless, M., Ellis, R. S., Taylor, K., \& Hook, R. N. 1990,
	\mn {244}, 408}
%\hi{Couch, W. A.} %NEW
\hi{Cowie, L. L., Songaila, A., \& Hu, M. 1991, {\it Nature}, {354}, 460}
\hi{Davis, M. \& Peebles, P. J. E. 1983, \apj 267, 465}
\hi{Dressler, A., \& Gunn, J. E. 1983, \apj 270, 1}
\hi{Eales, S. 1993, \apj 404, 51}
\hi{Efstathiou, G., Bernstein, G., Katz, N., Tyson, J. A.,
        \& Guhathakurta, P. 1991, \apj 380, L47}
\hi{Infante, L., and Pritchet, C.J. 1992, \apjs 83, 237}
\hi{Kron, R. G. 1980, \apjs 43, 305}
\hi{Lake, G. \& Tremaine, S. D. 1980, \apjl 238, L13}
\hi{Larson, R. B. \& Tinsley, B. M. 1978, \apj 219, 46}
\hi{Lavery, R. J. , \& Henry, J. P. 1988, \apj  330, 596}
\hi{Lavery, R. J., Henry, J. P., \& McClure, R. D. 1992, \aj 104, 2067}
\hi{Lilly, S. J., Cowie, L. L., and Gardner, J. P. 1991, \apj {\bf 369}, 79}
\hi{Lonsdale, C. J., Hacking, P. B., Conrow, T. P.,
         \& Rowan-Robinson, M. 1990, \apj  358, 60}
\hi{Maddox, S. J., Efstathiou, G., Sutherland, W. J. and Loveday, J. 1990,
	\mn 242, 43P}
\hi{Nilson, P. 1973, {\it Uppsala General Catalogue of Galaxies},
	(Uppsala: Royal Society of Sciences of Uppsala)}
%%%\hi{Peebles, P. J. E. 1980, {\it The Large-Scale Structure of the
%%%	Universe}, Princeton University Press}
\hi{Press, W. H. \& Schechter, P. 1974, \apj 187, 425}
\hi{Pritchet, C. J. \& Infante, L. 1992, \apjl {399}, L35}
\hi{Rocca-Volmerange, B. \& Guiderdoni, B. 1990, \mn 247, 166}
\hi{Steidel, C. 1993, in ``The Environment and Evolution of Galaxies'',
	proc. 3rd Teton Conference, eds. J. M. Shull and H. A. Thronson,
	(Dordrecht: Kluwer), p. 263}
%%%\hi{Sharp, N. A. \& Jones, B. J. T. 1980, \nature 283, 275}
\hi{Toomre, A. 1977, in {\sl Evolution of Galaxies and
    Stellar Populations,} ed. B. M. Tinsley \& R. B. Larson,
    (Yale Observatory: New Haven) p. 401}
\hi{Wolfe, A. M., Turnshek, D. A., Smith, H. E., and Cohen
	R. D.  1986, ApJS 61, 249}
\hi{Zepf, S. E., and Koo, D. C. 1989, \apj 337, 34}

\ifapj
\vfill\eject
\medskip
\copy21
\bigskip\bigskip\bigskip
\copy22
\vfil\eject
\ni\ub{FIGURE CAPTIONS}
\ls
\medskip
\ifpairpics\copy11\bigskip\fi
\copy12\vfil\eject
\ifpairpics
\pageinsert\copy31\endinsert
\pageinsert\copy32\endinsert\fi
\centerline{\caps Fig. 1}
\vfil
\epsfysize=8.5truein
\centerline{\epsffile{mg.ps}}
\fi

\bye